\documentstyle[aps,draft,epsf]{revtex}

\begin{document}

%
%

\title{Ultra High Energy Neutrinos from Supernova Remnants}
 \vspace{1.3cm}
\author{Mou Roy\\ }
\address{MS 50-245, Lawrence Berkeley National Laboratory, \\
         One Cyclotron Road, Berkeley, CA 94720.}
\preprint{}
\baselineskip 21 pt

\maketitle

\begin{abstract}

In this paper we discuss possible ultra high energy ( $\ge$ TeV) neutrino
emission from Supernova Remnants (SNRs), specifically the 
hadronic gamma ray production models.
Recent very high energy (VHE) $\gamma$ ray observation 
from SNRs is the main motivation
behind this study.

\end{abstract}
 
\bigskip\bigskip

\section{Introduction}

The supernova
remnants (SNRs) could be the principal source of galactic cosmic rays
up to energies of $\sim 10 ^{15}$ eV \cite{snr1}. A fraction of the accelerated
particles interact within the supernova  remnants and its adjacent
neighborhood, and produce
$\gamma $ rays.
If the nuclear component of cosmic rays is strongly enhanced inside
SNRs, then through nuclear collisions 
leading to pion production and subsequent decay, $\gamma$ rays and $\nu$s are
produced. Therefore, simultaneous high energy $\gamma$ ray and $\nu$ 
observations from SNR sources would suggest accelerated hadrons in SNR.

Recent observations above 100 MeV by the EGRET
instrument have found $\gamma$
ray emission from the direction of several SNRs 
(e.g. IC 443, $\gamma$ cygni, etc.).
However, the production mechanisms of these high energy gamma-rays
has not been unambiguously identified. The
emission may be due to the interaction of protons, accelerated by the
SNR blast wave, with adjacent molecular clouds \cite{drury},
bremsstrahlung or inverse Compton from accelerated electrons
\cite{JM}, or due to pulsars residing
within the SNRs.
Evidence for electron acceleration in SNR comes from the ASCA satellite
detection of non-thermal X-ray emission from SN 1006 \cite{koyama}
and IC 443 \cite{keohane}. Ground based telescopes have detected TeV emission
from SN 1006 \cite{cangre} and the Crab Nebula \cite{cangaroo}.
For recent reviews see \cite{snr1}.
Our objective in this paper is to look closely at SNRs as sources
of ultra high energy (UHE) neutrinos and we investigate different
possibilities in the sections II and III below. Section IV gives a
brief overview and conclusions.

\section{Neutrinos from Supernova Remnants Assuming p-p Interactions}

There are two schools of thought describing the high 
and very high energy gamma-ray emission from SNRs.
In one, TeV $\gamma $ rays are suggested to
be leptonic in origin \cite{leptonic} where 
TeV photons are produced in inverse Compton scattering off the
microwave radiation and other ambient photon fields by
relativistic electrons. 
In the second,  the decay of neutral pions produced in
proton nucleon collisions produce gamma-rays. 
TeV neutrino emission from SNRs is possible only if hadronic 
models are taken into consideration.

Non-linear particle acceleration concepts
have been used in \cite{drury} to provide SNR gamma-ray fluxes.
Following \cite{drury} 
the gamma-ray flux above 1 TeV from a SNR at a distance, d, 
and considering a differential energy spectrum of accelerated protons 
inside the remnant of the form $E^{2 - \alpha}$, would be

\begin{equation}
F_{\gamma}( >{1 \;\rm TeV}) \sim 8.4 \times 10^6 \;\; \theta \; q_\gamma (\alpha)
{\left ( {E \over 1 {\rm TeV}} \right )}^{3 - \alpha} \left ( {E_{\rm sn} \over
10^{51}{\rm erg}
} \right ) \left ( {n \over
1{\rm cm}^{-3}} \right ) {\left ( { {\rm d} \over 1{\rm kpc}} \right )}^{-2}
 \;\; {\rm cm}^{-2} {\rm s}^{-1}
\end{equation}

\noindent
where $n$ is the number density of the gas and $q_\gamma$ is the production
rate of photons. 
These results correspond to the SNR in the Sedov (adiabatic) phase where 
the luminosity is
roughly constant. 

UHE neutrinos can be predicted to be produced
as a significant byproduct of the 
decay of charged pions.
To find the neutrino flux ($F_{{\nu_\mu} + {\bar{\nu_\mu}}}$)
for different
spectral indices  we resort to the calculated ratios 
${F_{{\nu_\mu} + {\bar{\nu_\mu}}} \over F_\gamma} $ \cite{drury,gaisserbook} 
as given in Table 1.
$q_\gamma (\alpha)$ values for different  $\alpha$ 
are also included.
The contribution of nuclei other than H in both the target matter
and cosmic rays is assumed to be the same as in the ISM \cite{gamma}.
The units of $q_\gamma$ are 
${\rm s}^{-1}\; {\rm erg}^{-1}\; {\rm cm}^{3}\; {\rm (H)}^{-1}$.
A comprehensive discussion of the spectrum weighted moments for secondary
hadrons, based on the accelerator beams with fixed targets at beam energies
$\le 1$ TeV, has been presented by Gaisser \cite{gaisserbook}. This has been
shown to also characterize correctly the
energy region beyond 1 TeV \cite{drury}. A direct ratio estimate was
calculated in \cite{drury} to give results
very close to that in \cite{gaisserbook}. For harder spectra the ratio is found
to approach unity.

\vspace{0.5 cm}
\begin{table}[h]
\begin{center}
\begin{tabular} {|l|c|c|c|c|}
\vspace{0.25 cm}
       $\alpha$ & 4.2 & 4.4 & 4.6 & 4.8 \\
\hline
\vspace{0.25 cm}
${F_{{\nu_\mu} + {\bar{\nu_\mu}}} \over F_\gamma}$ 
[Gaisser (1990)\cite{gaisserbook}] &0.80&0.67&0.56&0.46 \\ \hline
\vspace{0.25 cm}
${F_{{\nu_\mu} + {\bar{\nu_\mu}}} \over F_\gamma}$ 
[Drury {\it et.al.} (1993)\cite{drury}]  &0.86&0.77&0.66&0.58 \\ \hline
\vspace{0.25 cm}
$q_\gamma (\alpha)$ \cite{drury} & $4.9 \times 10^{-18}$ & $8.1 
\times 10^{-19}$  &$1.0 \times 10^{-19}$ & $ 3.7 \times 10^{-20}$ \\
\end{tabular}
\vspace{0.5 cm}
\caption{Values of expected UHE neutrino and gamma ray ratio}
\end{center}
\end{table}

\nopagebreak

We have taken the average of the two ratios as given in
Table 1 to calculate the corresponding
neutrino flux from equation (1) for each spectral index.
The expression for the $\nu_\mu$  flux for
$\alpha \sim 4.2$ is

\begin{equation}
F_{\nu_\mu}( > {1 \; \rm TeV}) \sim 3.4 \times 10^{-11} \;\;  
\theta {\left ( {E \over 1 {\rm TeV}} \right )}^{-1.2}
\left ( {E_{\rm sn} \over 10^{51}{\rm erg}}\right )
\left ( {n \over 1{\rm cm}^{-3}} \right )
{\left ( { {\rm d} \over 1{\rm kpc}} \right )}^{-2}
 \;\; {\rm cm}^{-2} {\rm s}^{-1}
\end{equation}

\noindent
This is twice the  corresponding $\nu_e$ flux.  

Recent data from the CANGAROO detector \cite{cangaroo} indicate that the
energy spectrum of $\gamma$ rays from the Crab pulsar/nebula
may extend up to at least 50 TeV.
The CANGAROO detector has also observed VHE $\gamma$ rays
upto 10 TeV from SN1006 \cite{cangre}.
These emissions could be explained on the basis of  electron 
inverse Compton processes.
However, as the energy increases,
Compton process produces  steeper spectra because of
synchrotron energy loss of electrons in magnetic fields.
Hence, leptonic models have difficulty in explaining the observed 
hard spectrum that extends to
beyond 10 TeV as is observed from the Crab.
If we consider hadronic models to be viable at such energies,
we should expect corresponding UHE neutrino emission 
from SNRs.

The hadronic mechanism for production of $\pi^0$
and hence $\gamma$ rays as described above has been used in \cite{aha}
to explain the UHE emission from the Crab.
Nuclear p-p interactions are considered to occur among
protons accelerated in the nebula. 
However, 
the energy balance
between the magnetic field and relativistic particles in the
nebula show that nucleon contribution is dominant
 only at energies above 10 TeV.
The derived gamma-ray spectrum
in this model closely matches the SN1006 spectrum
obtained by the CANGAROO instrument.

\setbox4=\vbox to 180 pt{\epsfysize=5.5 truein\epsfbox[20 -220 632 572]{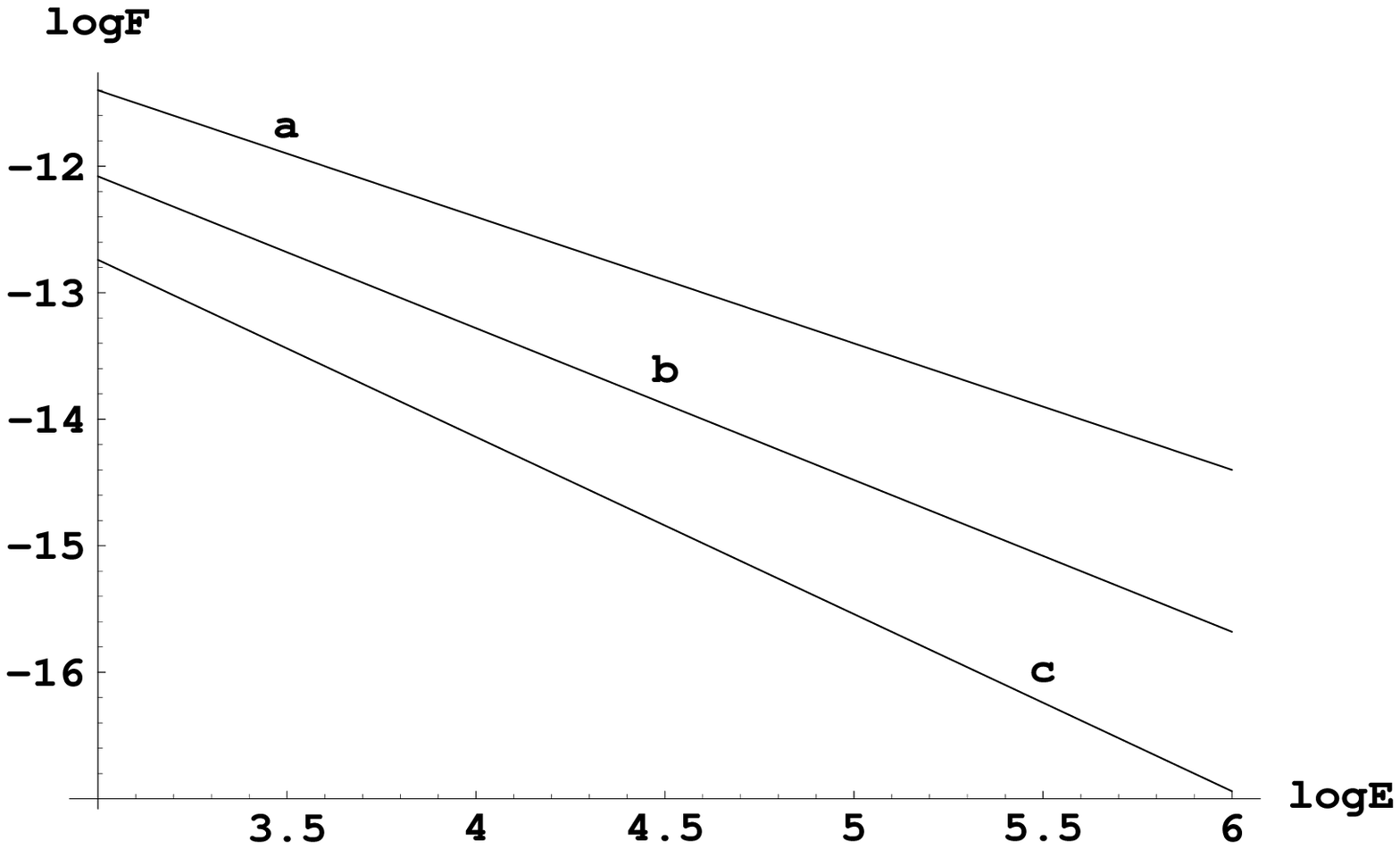}}
\setbox5=\vbox to 180 pt{\epsfysize=5.5 truein\epsfbox[20 -220 632 572]{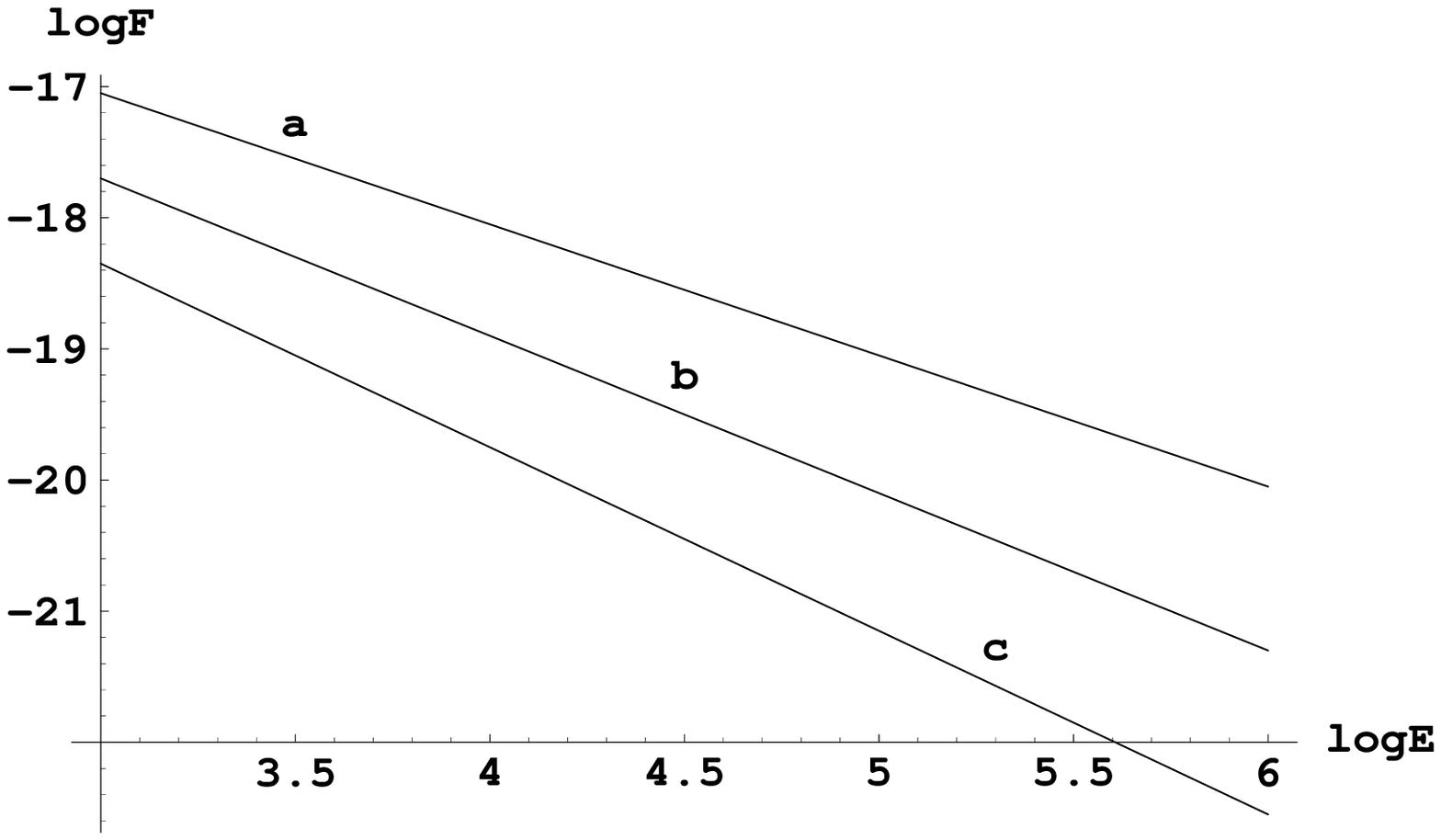}}
\begin{figure}
\centerline{\hfill \box4 \box5 \hfill}
\caption{Plot of expected neutrino flux from the Crab (left) and SN1006 (right)
considering
hard source spectra, $\alpha$ $\sim$ 4.0 (a), 4.2(b), 4.4(c).
Here F is in ${\rm cm}^{-2} {\rm s}^{-1}$ and
neutrino energy E is in GeV.
\label{fig:f.2}}
\end{figure}

An approximate expression for the gamma ray spectrum
from equation (1) could be written 
for spectral index $\alpha$
varying between 4.0-4.5 as \cite{aha},

\begin{equation}
F_{\gamma} ( > {1 \rm TeV}) \sim 4.0 \times 10^{- {3 \alpha}}\;\;
\left ( {W_p \over 10^{48}{\rm erg}} \right )
\left ( {n \over 100 {\rm cm}^{-3}} \right )
{\left ( { {\rm d} \over 2 {\rm kpc}} \right )}^{-2}
 {\left ( {E \over {\rm TeV}}
\right )}^{3 - \alpha} \;\; {\rm cm}^{-2} {\rm s}^{-1}
\end{equation}

\noindent
where d is the distance to the source (distance to Crab is 2 kpc),
$W_p$ is the kinetic energy of the accelerated protons (reasonable
value is $\sim 10^{48}$ erg) and $n$ is the effective number density
(100 ${\rm cm}^{-3}$).
The corresponding UHE neutrino flux can be calculated directly
using values from Table 1. For reasonable parameters as an example
for $\alpha \sim 4.2$ the neutrino flux from the Crab would be,

\begin{equation}
F_{\nu} ( > { 1 \rm TeV}) \sim 8.3 \times 10^{-13}
{\left ( {E \over {\rm TeV}} \right )}^{-1.2} \;\; {\rm cm}^{-2} {\rm s}^{-1}
\end{equation}

\noindent
In the shell type supernova remnant, SN1006, 
at a  distance of $\sim 1.8$ Kpc 
\cite{leptonic},
the total estimated kinetic energy of the  accelerated protons would be
$\sim 10^{49}$ ergs ($10\%$ of the supernova explosion energy of 
$\sim 10^{50}$ ergs \cite{masta2}) and the matter density is 
$0.4 \; {\rm cm}^{-3}$ \cite{willi} (The matter density is low since
 SN1006 is located 
above the galactic plane). The
observed $\gamma$ ray flux cannot be accounted for by the hadronic acceleration
mechanism alone due to this low matter density. However, there could be 
corresponding UHE neutrino emission.

We show in Figure 1 the  expected neutrino flux from
the Crab Nebula and SN1006 for different spectral indices as calculated above
for typical parametric constants. The expected neutrino
flux for SN1006 is found to be
several orders of magnitude lower than the Crab for the same neutrino
energies.

\section{UHE Neutrino Emission from SNR and Pulsars due to Nuclear Interactions}

There is another model which predicts UHE neutrinos from SNR
\cite{protheroesnr,protheroepuls}.
In this model, very young SNRs are considered in which ions are
accelerated in the slot gap of the highly magnetized
rapidly spinning pulsar.
Nuclei, probably mainly Fe nuclei, extracted from the neutron
star surface and accelerated to high Lorentz factors can be
photodisintegrated by interaction with  neutron star's
radiation field and hot polar caps. Photodisintegration can also occur in the
presence of extremely
strong magnetic fields typical of neutron star environments ($\sim 10^{12}$ G).
For acceleration to sufficiently high energies we need a short
initial pulsar period ($\sim$ 5 ms).
The energetic neutrons produced as a result of photodisintegration
interact with target
nuclei (matter in the shell) as they travel out of the SNR, producing
 gamma ray
and neutrino signals; those neutrons 
passing through the shell decay into relativistic
protons contributing to the pool of galactic cosmic rays.
For a beaming solid angle to the Earth of $\Omega_b$, 
the neutrino flux in this model can be calculated from,

\begin{equation}
F_\nu(E_\nu) \sim  { {\dot N}_{\rm Fe} \over {\Omega_b d^2}} [ 1 - {\rm exp}(
-\tau_{pp})] \int N_n (E_n) P_{n\nu}^M (E_{\nu},E_n) d E_n
\end{equation}

\noindent
where ${\dot N}_{\rm Fe}$ is the total rate of Fe nuclei injected,
$d$ is the distance to the SNR, $P_{n\nu}^M (E_{\nu},E_n) d E_n$ is the
number of neutrinos produced with energies in the range $E_{\nu}$ to
$(E_{\nu}+ d E_{\nu})$ (via pion production and subsequent decay),
and $ N_n (E_n)$ is the spectrum of neutrons extracted
from a single Fe nucleus. $\tau_{pp}$ is the optical depth of the
shell to nuclear collisions
(assuming shell type SNR) which is a function of the mass ejected into the
shell during the supernova explosion and of the time after explosion.
We show in Figure 2 the $\nu_\mu + {\bar \nu_\mu}$ spectra
obtained from this model at a distance of 10 kpc; the time
after explosion is 0.1 year.
Signals from nuclei, which are not completely fragmented, are ignored.
These particles are charged and would be trapped in the central
region of the SNR which has a relatively low matter density
and therefore would not make any significant contribution to
neutrino fluxes.

\setbox4=\vbox to 160 pt {\epsfysize= 5 truein\epsfbox[0 -200 612 592]{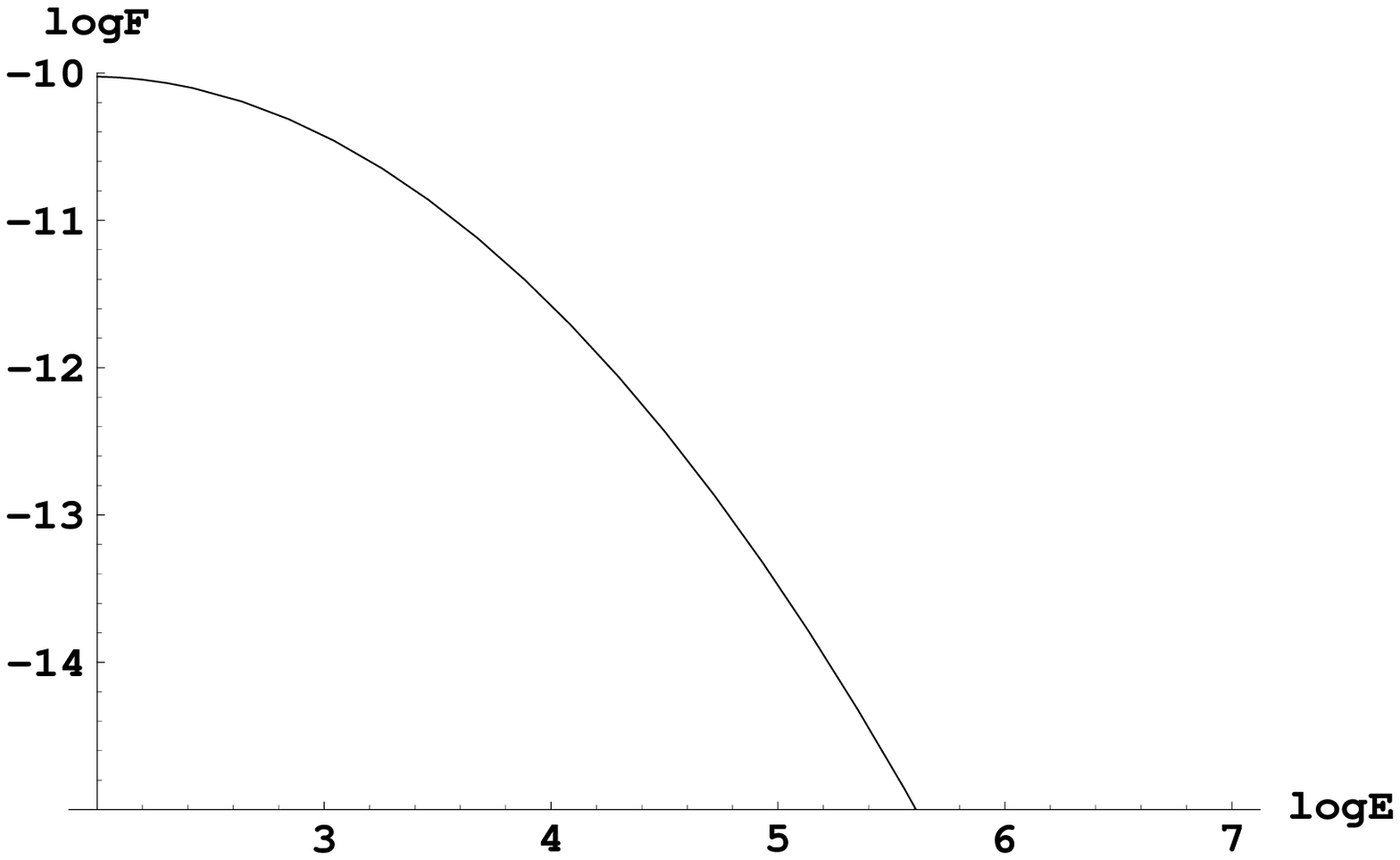}}
\begin{figure}
\centerline{\box4}
\caption{Plot of expected neutrino $ \nu_\mu + {\bar \nu_\mu} $
spectrum from SNR produced by collisions of neutrons with matter in a
supernova shell for B = $10^{12}$ G and initial pulsar period
5 ms using the maximim polar cap heating
model (very young SNR);
Here F is in ${\rm cm}^{-2} \;\; {\rm s}^{-1}$  and neutrino energy E is in GeV.
\label{fig:f.3}}
\end{figure}

\section{Overview and Conclusions}

UHE neutrinos can be detected by observing muons, electrons
and tau leptons produced
in charged-current neutrino nucleon interactions \cite{raj,pakvasa}.
To minimize the effects of atmospheric muon and neutrino background,
usually the upward going muons (to identify muon neutrinos) are observed.
To observe $\nu_e$, one looks at the contained event rates for
resonant formation of
$ W^-$ in the $ \bar{\nu_e}$ interactions at $E_{\nu} = 6.3$ PeV
for downward moving $\nu_e$.
The key signature for the detection of $\nu_\tau$ is the charged current
$\nu_\tau$ interaction, which produces a double cascade on either end
of a minimum ionizing track \cite{pakvasa}.
The threshold energy for detecting these 
neutrinos is near 1 PeV.
At this energy cascades are separated by roughly 100m which should be
resolvable in the planned neutrino telescopes.
However, the evidence for $\nu_\tau$ would indicate neutrino oscillations
since they are not expected from the hadronic models. 
Neutrinos produced by cosmic ray
interactions in the atmosphere are considerably larger than individual
source fluxes at 1 TeV but falls rapidly with energy.
The ``conventional" atmospheric neutrino flux
is derived from the decay of
charged pions and kaons produced by cosmic ray interactions in the atmosphere.
The angle averaged atmospheric flux in the neutrino energy range
1 TeV $< {E_\nu} < 10^3$ TeV, can be parametrized 
\cite{raj} by the equation

\begin{equation}
F_\nu = 7.8 \times 10^{-8}{\left( {E_\nu \over 1 {\rm TeV}} \right)}^{-2.6}
{\rm cm}^{-2} \;\; {\rm s}^{-1} \; {\rm sr}^{-1}
\end{equation}

\noindent
An additional ``prompt" contribution of neutrinos
to the atmospheric flux arises from charm production and decay.
The vertical prompt neutrino flux has been recently reexamined using the
Lund model for particle production \cite{lund} and has been shown to be 
slightly larger than the conventional atmospheric neutrino flux at
higher energies $\ge 100 $ TeV.
We compare equations (3) and (5) to find that the neutrino flux
from the Crab would be significantly above the atmospheric background
beyond a few TeV.
For energies of a TeV or more the neutrino
direction can be reconstructed to better than 1 degree,
and less than one event per year in a one degree bin
is expected from the combined atmospheric
and AGN backgrounds \cite{gaisser,lipari}.
For a muon
neutrino of energy $\sim$ 1 TeV
the rate of upward muons in a detector with effective area of $1 {\rm km}^2$
from the Crab
will be $\sim$ 1 -- 30 per year, depending on the model chosen.
This neutrino flux
should be detectable by large neutrino telescopes with
good angular resolution of about 1 degree.
However, neutrino flux from SN1006 will be negligible even in such large
area detectors.

We must also account for the shadow factor which represents the
attenuation of the neutrinos traversing the earth.
This effect is prominant at energies $\ge$ 100 TeV. 
In that case, it is necessary to
restrict our attention to downward moving neutrinos. 
The expected rates would
be larger, but the effects of atmospheric muons have to be eliminated
by restricting the solid angle to include only large column depths \cite{raj}.

The question of the importance of hadronic interactions in the Crab
Nebula and other SNRs can therefore 
be settled by the detection of  neutrinos 
which is likely in the next generation UHE kilometer scale detectors in
ice/water.

Acknowledgements : M.R. wishes to acknowledge useful discussions with
Dr. H.J. Crawford and Dr. D. Bhattacharya. Thanks to an anonymous referee for
useful comments. This research was supported in part by Grant NAG5-5146.

\end{document}